\documentclass{article}

\usepackage[dblblindworkshop, final]{neurips_2026}

\usepackage[utf8]{inputenc} 
\usepackage[T1]{fontenc}    
\usepackage{hyperref}       
\usepackage{url}            
\usepackage{booktabs}       
\usepackage{amsfonts}       
\usepackage{nicefrac}       
\usepackage{microtype}      
\usepackage{xcolor}         
\usepackage{graphicx}
\usepackage{fvextra}
\usepackage{tabularx}
\usepackage{adjustbox}

\hypersetup{hidelinks}

\usepackage{enumitem}

\DefineVerbatimEnvironment{Prompt}{Verbatim}{
  breaklines=true,
  breakanywhere=true,
  fontsize=\footnotesize,
  frame=single,
  framesep=2mm,
  baselinestretch=0.84
}

\setcitestyle{square, semicolon}

\title{A Framework for Context Enrichment and Selective Robustness in LLM Social Agents}
\workshoptitle{SocialAgent: Second Workshop on Large Language Models for Social Reasoning and Simulation}

\author{
  Mattia Greiche$^{1,2}$\thanks{Corresponding author: mattia.greiche@mail.mcgill.ca} \And
  Reihaneh Rabbany$^{1,2}$\And
  Maximilian Puelma Touzel$^{2}$\and \\[0.5em]
  $^1$McGill University, \quad
  $^2$Mila - Quebec Artificial Intelligence Institute
}

\begin{document}

\maketitle

\begin{abstract}

Researchers use large language models as agents in social simulations, where agents are prompted with observations and background information intended to shape their actions. Yet LLM agents have been shown to substantially change their behavior under meaning-preserving prompt variations, creating a validation problem for social simulation. One plausible hypothesis is that this brittleness reflects a lack of grounding and can be reduced by enriching the context provided to the agents. Whether richer context improves robustness remains unclear, in part because context richness and robustness are rarely defined in ways that support controlled comparisons. We address the first need with \emph{controlled context enrichment}, a set of prompt substitution rules that systematically enriches distinct types of context: neutral padding, generic social background, descriptive decision grounding, and normative decision grounding. We address the second by introducing \emph{selective robustness} as a validation criterion: insensitivity to meaning-preserving prompt variation combined with a directional response to decision-relevant changes in the social world. We demonstrate the framework in a Prisoner's Dilemma-like game, showing how it can be used to identify which forms of context enrichment produce more desirable and robust agent behavior. We find that these forms of enrichment produce distinct behavioral patterns that vary between the two models tested.

\end{abstract}

\section{Introduction}
Researchers use large language models (LLMs) as agents in simulations of cooperation and other social processes~\citep{mou2026survey}. Generally, an agent is provided context through a role and a setting in which it acts through natural language; they may also be given information related to relationships, biography, memories, goals, or beliefs. Concordia uses such descriptions to ground agent actions in social or physical spaces~\citep{vezhnevets2023concordia}. Similarly, SOTOPIA assigns characters and social goals within interactive scenarios~\citep{zhou2024sotopia}. Researchers then interpret the agent's choices or actions as behavior within the supplied social world. The context provided to the agents, therefore, is a fundamental variable in any LLM agent simulation's setup.

Recent reviews and position papers identify validation as a central unresolved problem for LLM social simulation and call for more standardized, reproducible evaluation systems~\citep{larooij2026validation,puelmatouzel2026validationgap}. One challenge is that small changes in how researchers present the same social information can alter agent behavior, including changes to wording and prompt formatting~\citep{ju2024sense, lutz2025prompt}. Recent work by \citet{ye2026trails} reports shifts of up to 76 percentage points in cooperation rates in a Prisoner's Dilemma game when changes are made to context format that maintain similar social meaning. Such findings make stability under meaning-preserving changes an important validation test. But universal stability does not show that an agent uses the social context appropriately; ideally, an agent \emph{should} respond and alter its behavior when the social world changes in a way that bears on the decision.

We call this joint requirement \emph{selective robustness}: ideally, an agent should be insensitive to changes in representation that preserve the social meaning relevant to the decision, but sensitive to substantive changes in the social world that likely should affect the decision. This framing follows the broader logic of metamorphic testing~\citep{chen2018metamorphic}, and is inspired by CheckList's behavioral testing methodology, which distinguishes invariance tests, where small perturbations to the input should leave predictions unchanged, from directional expectation tests, where a controlled input change should change predictions in a specified way~\citep{ribeiro2020checklist}. We adapt this to the LLM social agent domain by testing both stability on different representations of the same social world and responsiveness to a prespecified change in that world.

Prompting with richer context has been proposed as one route to more accurate and diverse social simulation~\citep{anthis2025promising}. More detailed persona profiles have also brought greater stability across repeated tests~\citep{bai2025personality}. However, context enrichment is not a single well-defined intervention. Additional context may simply add text, provide generic social information, connect the agent more directly to the current decision, or introduce explicit normative guidance. Treating these changes only as increases in context ``richness'' makes it difficult to determine which property is responsible for any changes in behavior. We therefore decompose context enrichment according to the function of the added information, allowing these effects to be compared more systematically.

\paragraph{Our approach.}
We make three contributions:
\begin{itemize}
    \item \textit{a controlled decomposition of context enrichment across five prompt conditions} made up of one minimal-context baseline and four enrichment conditions: neutral padding, generic social background, descriptive decision grounding, and normative decision grounding;
    \item \textit{selective robustness as an evaluation criterion} that combines surface invariance with responsiveness to meaningful changes in the social world; and
    \item \textit{a grounded pilot study} showing how these two ideas can reveal different behavioral effects of context construction and different failure modes across two models.
\end{itemize}

The pilot study is run in only one global scenario. Its results are intended to illustrate the framework's utility, not to establish a general effect of context enrichment in social simulations. The codebase and data for the pilot experiment are
available at \url{https://github.com/mattiagreiche/context-enrichment-selective-robustness}.

\section{Framework}

\subsection{Controlled context enrichment}
\label{sec:context-enrichment}
We operationalize context enrichment as a rule-based set of substitutions within a common prompt scaffold. Across all five prompt conditions, we fix the focal role and decision task and use the same substantive social-world contrast: a prespecified and meaningful change in the world for which agent behavior should shift in a specified direction. Adding a social-world contrast allows us to check whether the agent responds to meaningful changes in the world, as selective robustness requires. 

The minimal context condition has no enrichment. Each of the other four conditions inserts one context block at the same location, with the blocks replacing rather than accumulating on one another. We order the conditions for comparison according to how the added information relates to the agent and the current decision. This ordering distinguishes changes in function and does not serve as a gradient of longer nor necessarily ``richer'' information.

\begin{table}[t]
\centering
\caption{\textit{Functions of the five context conditions.} The role of the agent, the decision task, and the decision-relevant world contrast are always present. The enrichment (non-minimal) blocks substitute for one another rather than accumulate.}
\setlength{\tabcolsep}{3pt}
\begin{tabular*}{\linewidth}{@{\extracolsep{\fill}}llccc@{}}
\toprule
Condition & Added context
  & \shortstack{World\\contrast} & \shortstack{Decision\\link} & \shortstack{Normative\\force} \\
\midrule
1. Minimal               & None                               & $\checkmark$ & --           & --           \\
2. Neutral padding       & Unrelated background detail        & $\checkmark$ & --           & --           \\
3. Generic social background        & Roles not tied to the decision     & $\checkmark$ & --           & --           \\
4. Descriptive decision grounding & Relationships, rule, applicability & $\checkmark$ & $\checkmark$ & --           \\
5. Normative decision grounding   & Same as 4                          & $\checkmark$ & $\checkmark$ & $\checkmark$ \\
\bottomrule
\end{tabular*}
\label{tab:context-conditions}
\end{table}

\begin{enumerate}
    \item \textbf{Minimal: task and world specification.} 
This contains only the information needed to identify the agent's role, understand the available decision, and distinguish the relevant social worlds.

\item \textbf{Neutral padding: additional background detail without agent grounding.}
This adds background information that should not extend the agent's social position or connect to the decision.

\item \textbf{Generic social background: social information but not decision-relevant.}
This adds information about the agent's role within an organization or community, such as relationships, group membership, or any recurring social roles. This information better situates the agent socially, but is still disconnected from the current decision and its procedure.

\item \textbf{Descriptive decision grounding: relationship, procedure, applicability.}
This connects the agent and relevant counterpart(s) to the social structure of the current decision. It specifies relevant relationships, the scope of a procedure or rule, and how that applies to the decision. The added information is descriptive, i.e. it does not assign an obligation nor recommend an action. 

\item \textbf{Normative decision grounding: directed role requirement.} 
This keeps the relational and procedural facts from descriptive decision grounding, but expresses requirements or responsibilities explicitly. It introduces normative force but does not go so far as to prescribe an action.
\end{enumerate}

\subsection{Selective robustness}
\label{sec:selective-robustness}
Robustness should be selective: surface-level changes to a prompt should leave behavior stable, while substantively meaningful changes should alter behavior. We elaborate on these two properties below.

\paragraph{Surface invariance.}
Within a fixed world and context condition, behavior should remain similar across alternative representations or formats intended to preserve decision-relevant meaning. Prior robustness audits have tested this type of sensitivity by varying persona format, game framing, and narratives~\citep{ju2024sense, lutz2025prompt, ye2026trails}. Large differences across representations would indicate sensitivity to surface form, rather than to the underlying social situation.

\paragraph{World responsiveness.}
Within a fixed context condition and representation, behavior should respond to a change in decision-relevant social information. Examples include changing the relationship between actors or institutional rules governing the choice, while holding the available actions and material payoffs fixed. The expected direction of the change should be prespecified, i.e. determined before observing the model's behavior.

Neither property is sufficient on its own. Low surface sensitivity can be a sign of robustness, but it can also arise if the model simply ignores the social context or consistently chooses the same action. Conversely, strong responsiveness to the social world may be overtly brittle if it depends heavily on how that world is represented.

\section{Experiment}

We demonstrate the framework in a pilot experiment within SiliSocS~\citep{sarangi2026silisocsdemo}, a configurable system for social simulation. We prompt one focal agent, which acts as the emergency-supply manager for Hospital A, in a symmetric Prisoner's-Dilemma-like decision. The focal agent privately chooses whether to \textsc{transfer} an emergency supply batch to the other hospital or to \textsc{retain} it, and is told that the manager of Hospital B faces the same choice. We use two worlds with a decision-relevant distinction: one world has the hospitals governed by a shared regional arrangement, and the other by separate local arrangements. We expected more transfer in the regional world because its emergency-supply arrangement covers the supply needs of both hospitals. We cross the two worlds with the five context constructions in Section~\ref{sec:context-enrichment} and four surface forms intended to preserve meaning: canonical prose, bullets, a key-value table, and alternative prose. We test GPT-5.6 Sol and Terra. Appendix~\ref{app:experiment-details} presents the game, prompts, and complete factorial design.

\paragraph{Analysis.}

For each call, we get the probabilities of \textsc{transfer} and \textsc{retain} from the model's token probabilities. We summarize each context with the two selective robustness measures from Section~\ref{sec:selective-robustness}: mean sensitivity to the surface forms across the two worlds (lower means more stable behavior) and regional-minus-local world responsiveness (positive values follow the prespecified direction). Appendix~\ref{app:experiment-measurement} describes probability extraction and aggregation.

\paragraph{Results.}

\begin{figure}[t]
  \centering
  \includegraphics[width=1\textwidth]{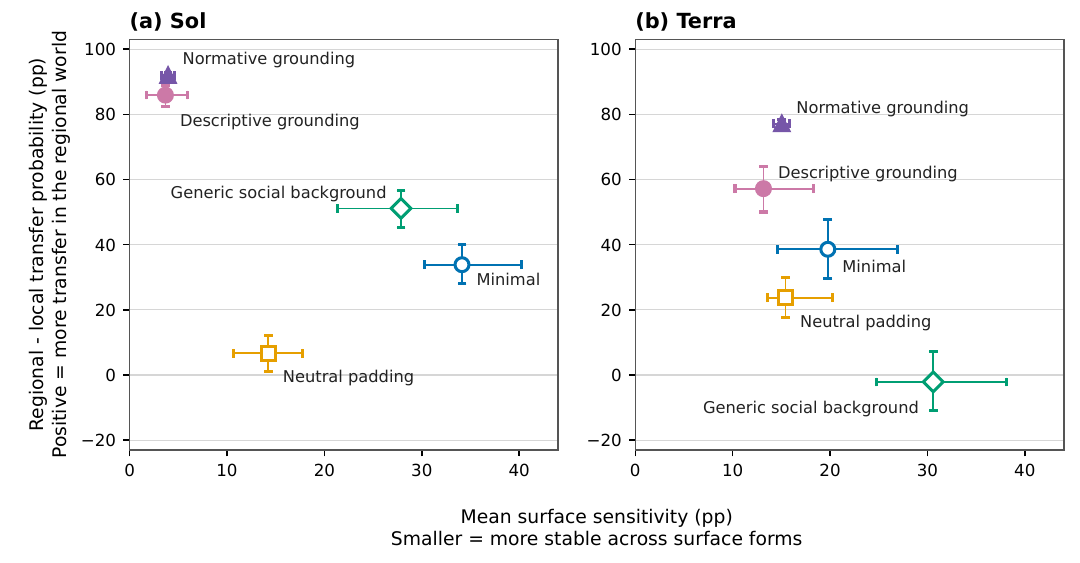}
  \caption{\textit{Selective robustness across five context constructions for (a) Sol and (b) Terra.} Each labeled point represents one prompt condition. The x-axis shows mean surface sensitivity: the average absolute difference in transfer probability between canonical prose and each of the three alternative surface forms, across both worlds. The y-axis shows the regional-minus-local difference in transfer probability averaged across surface forms; positive values indicate more transfer in the regional world. Horizontal and vertical bars show 95\% bootstrap intervals. Descriptive and normative grounding produce stronger world responsiveness for both models, but Sol combines this response with lower surface sensitivity than Terra.}
  \label{fig:experiment-results}
\end{figure}

Figure~\ref{fig:experiment-results} shows the main selective robustness comparison. The non-grounded conditions generally fail to produce the desired combination of surface stability and world responsiveness. Some are more sensitive to surface form, while others are relatively stable but weakly respond to the world distinction. 

Under descriptive and normative grounding, both models become strongly responsive to the world distinction. Sol also becomes more stable across surface forms, placing it close to the preferred selective-robustness region (upper left). Terra remains sensitive to surface form despite responding strongly to the world distinction. Because the grounded prompts also elaborate on the world distinction, their stronger world response could reflect greater salience of that information as well as decision grounding.

In Terra's local normative condition, alternative prose yields 89.9\% transfer, compared with at most 0.9\% for the other three surfaces. This large shift under a surface variation intended to preserve meaning echoes the findings of \citet{ye2026trails}. In the same vein, the different trajectories of Sol and Terra are reminiscent of prior studies that find sensitivity to the same simulation perturbation can vary substantially across models~\citep{sclar2024quantifying}. Appendix~\ref{app:experiment-additional-results} reports full results with added discussion.

\section{Discussion and Outlook}

We present a framework for controlled experiments on context enrichment in LLM social agents, motivated by the growing need to validate LLM-based social simulations. The framework uses substitutions to compare five context conditions: minimal, neutral padding, generic social background, descriptive decision grounding, and normative decision grounding. We also introduce selective robustness as an evaluation criterion, which calls for both stability under meaning-preserving surface changes and a directional response to meaningful changes in the social world. Our pilot experiment illustrates the potential value of both ideas: the five context conditions produced distinct patterns of surface stability and world responsiveness.

An interesting feature of the pilot is that the numerical payoffs are identical in the regional and local worlds, yet the grounded conditions produce large differences in transfer probability. This suggests that the models respond to social and institutional information beyond what the payoff table represents. Related work has found that social contexts can affect LLM play under fixed payoffs~\citep{lore2024strategic}. The models may draw on learned associations about institutional roles and responsibilities, although our experiment cannot establish how they arrive at their choices.

Researchers who develop social agent studies should treat context construction as part of their experimental design. Prompt format, the amount of context, and other design choices have been shown to alter outcomes with effects that vary across models~\citep{sclar2024quantifying, cummins2025analytic, buckkaeffer2026cookbook}. Thus, researchers should specify what each part of the supplied context is meant to do and use controlled substitutions to study effects. Robustness audits should include both surface perturbations and a social-world contrast to test for selective robustness.

Selective robustness does not by itself establish fidelity to human behavior; it serves as an internal diagnostic. A larger difference in transfer probability between the regional and local worlds shows stronger responsiveness in the expected direction, but does not tell us whether the size of that response is appropriate. One could imagine that humans are only moderately more likely to transfer in the regional world. In that case, a model with transfer probabilities of 100\% there and 0\% in the local world would follow the expected direction but exaggerate the size in comparison to the human response.

Because our pilot experiment covers only one decision setting and two models from the same family, the observed patterns—such as generally greater selective robustness in the grounded conditions—require further investigation before they can be generalized. Although we reviewed all prompts, no independent human raters assessed the semantic equivalence of the four surface forms or checked whether the non-normative prompts contained any unintended normative force. A potential confound is that the grounded prompt conditions add more detail about the social-world distinction than the non-grounded conditions, so the experiment does not fully separate the effect of decision grounding from increased salience of that distinction. Part of the difficulty in addressing this confound is that any plausible grounded prompt may naturally draw more attention to the relevant world information.

Future work could replicate the pilot across different decisions, world settings, and model families. Human raters could check the semantic equivalence of surface forms and detect any unintended normative force. Human decisions in the same settings could also help assess what magnitude of world responsiveness is plausible. A further experiment could compare decision grounding with other plausible ways of making the world distinction salient. Existing or future simulation frameworks could add prompt surface and world checks as reusable evaluation components to test for selective robustness.

\bibliography{references}


\appendix

\section{AI use statement}

LLM-based coding agents implemented the study-specific code under the authors' direction. We also used LLMs during manuscript preparation to discuss the paper's structure and arguments, suggest revisions, and draft passages for author review and revision. The authors made the final decisions about the study design, analyses, interpretation, and wording, and take full responsibility for the code, results, and paper.

\section{Experiment materials, measurement, and additional results}
\label{app:experiment-details}

\subsection{Game and factorial design}
\label{app:experiment-design}

The focal agent chooses whether to \textsc{transfer} Hospital A's designated emergency-supply batch to Hospital B or \textsc{retain} it. The focal agent is told that the other manager faces the same simultaneous choice. The fixed payoff table uses $T=5>R=3>P=1>S=0$, common in Prisoner's Dilemma games. Section~\ref{app:experiment-prompts} reproduces the complete prompts given to the agents.

The regional world places both hospitals under a shared arrangement covering supply continuity across them. The local world gives each hospital a separate arrangement within that hospital. We prespecified (expected) a higher transfer probability in the regional world. In all, crossing the two worlds with five context constructions and four surface versions results in $2 \times 5 \times 4 = 40$ cells per model.

The context constructions are minimal context, neutral padding, generic social background, descriptive decision grounding, and normative decision grounding. The surfaces are canonical prose, bullets, a key-value table, and alternative prose. We queried \texttt{gpt-5.6-sol} and \texttt{gpt-5.6-terra} ten times per cell, for 400 calls per model and 800 calls total. Every request had no conversational history.

\subsection{Prompt materials}
\label{app:experiment-prompts}

Each frozen prompt consists of a fixed opening, a social-context block in one of four surface formats, and a fixed decision block. We present the wording verbatim below in a modular form so that all 40 prompts can be reconstructed by repeating the shared text. Arrow glyphs in the displayed blocks indicate line wrapping and were not part of the prompt text.

We used \texttt{gpt-5.6-sol} to assist in drafting the prompt content and surface variations. The authors reviewed and finalized every prompt before execution.

\paragraph{Fixed opening.}

\begin{Prompt}
Hospital supply decision

Role

You are the emergency-supply manager for Hospital A.

Social context
\end{Prompt}

\paragraph{Fixed decision block.}

\begin{Prompt}
Decision task

Hospital A and Hospital B each control a designated batch of emergency supplies. You control Hospital A's batch, and the other manager controls Hospital B's batch. The action labels apply in the same way to both managers. For either manager, TRANSFER sends the designated batch to the other hospital, while RETAIN keeps the designated batch at the hospital that controls it.

You and the other manager choose at the same time. You cannot communicate, and neither of you sees the other's choice before deciding. This decision concerns the current allocation. The hospitals have an ongoing operational relationship, and similar allocation situations may arise in the future.

Actions

- TRANSFER: You send Hospital A's designated batch to Hospital B.
- RETAIN: You keep Hospital A's designated batch at Hospital A.

Payoff table

The batches contain different supplies. Each hospital gains a supply benefit from receiving the other hospital's batch and incurs a handling cost when sending its own. The table applies these effects symmetrically and reports outcome points for Hospital A and Hospital B.

Higher outcome points represent better outcomes for that hospital.

Your action | Other manager's action | Hospital A points | Hospital B points
--- | --- | ---: | ---:
TRANSFER | TRANSFER | 3 | 3
TRANSFER | RETAIN | 0 | 5
RETAIN | TRANSFER | 5 | 0
RETAIN | RETAIN | 1 | 1

Choose the action you would take in this role.

Respond with exactly one word: TRANSFER or RETAIN. Do not provide an explanation.
\end{Prompt}

\paragraph{Social-context block.}
In canonical prose, facts are separated by blank lines. The bullet format prefixes each fact with a hyphen. The key-value format contains one field-value row per fact. The world fact uses ``Institutional scope''; neutral padding uses ``Inventory system,'' ``Recordkeeping,'' and ``Routine schedule''; generic social background uses ``Team,'' ``Reporting line,'' and ``Organizational routine.'' Both grounded conditions use ``Counterpart relationship'' and ``Procedure,'' followed by ``Decision coverage'' for descriptive grounding or ``Role requirement'' for normative grounding. Alternative prose uses the corresponding paraphrases, separated by blank lines.

\paragraph{Canonical world facts.}

\noindent\textit{Regional world.}
\begin{Prompt}
For this allocation, the shared regional emergency-supply arrangement covers supply continuity across Hospital A and Hospital B.
\end{Prompt}

\noindent\textit{Local world.}
\begin{Prompt}
For this allocation, each hospital's separate local emergency-reserve arrangement covers supply continuity within that hospital.
\end{Prompt}

\paragraph{Canonical added modules.}

\noindent\textit{Neutral padding.}
\begin{Prompt}
Your supply office uses a standard electronic inventory system for routine stock records.

The office logs completed inventory updates on its usual daily recordkeeping form.

Staff review inventory totals during the supply office's scheduled afternoon meeting.
\end{Prompt}

\noindent\textit{Generic social background.}
\begin{Prompt}
You supervise two inventory assistants who work in the emergency-supply office.

You report to the operations director through the hospital's standard management structure.

You attend the hospital's weekly administrative planning meetings with managers from other departments.
\end{Prompt}

\noindent\textit{Regional descriptive grounding.}
\begin{Prompt}
You and the other manager are the designated supply contacts for two hospitals in the same regional emergency-supply network.

During shortages, the network's procedure assesses supply conditions across its participating hospitals.

The regional network's shortage procedure covers your decision about how to manage Hospital A's designated batch.
\end{Prompt}

\noindent\textit{Local descriptive grounding.}
\begin{Prompt}
You and the other manager are the designated supply contacts for hospitals in separate municipal emergency-supply systems.

During shortages, each system's procedure assesses supply conditions within its own hospital.

Hospital A's local shortage procedure covers your decision about how to manage Hospital A's designated batch.
\end{Prompt}

\noindent\textit{Regional normative grounding.}
\begin{Prompt}
You and the other manager are the designated supply contacts for two hospitals in the same regional emergency-supply network.

During shortages, the network's procedure assesses supply conditions across its participating hospitals.

Your role requires you to follow the regional network's shortage procedure when deciding how to manage Hospital A's designated batch.
\end{Prompt}

\noindent\textit{Local normative grounding.}
\begin{Prompt}
You and the other manager are the designated supply contacts for hospitals in separate municipal emergency-supply systems.

During shortages, each system's procedure assesses supply conditions within its own hospital.

Your role requires you to follow Hospital A's local shortage procedure when deciding how to manage Hospital A's designated batch.
\end{Prompt}

\paragraph{Alternative-prose world facts.}

\noindent\textit{Regional world.}
\begin{Prompt}
For this allocation, supply continuity across Hospital A and Hospital B is covered by the shared regional emergency-supply arrangement.
\end{Prompt}

\noindent\textit{Local world.}
\begin{Prompt}
For this allocation, supply continuity within each hospital is covered by that hospital's separate local emergency-reserve arrangement.
\end{Prompt}

\paragraph{Alternative-prose added modules.}

\noindent\textit{Neutral padding.}
\begin{Prompt}
Your supply office keeps routine stock records in a standard electronic inventory system.

The office uses its usual daily recordkeeping form to log completed inventory updates.

The supply office's scheduled afternoon meeting includes staff review of inventory totals.
\end{Prompt}

\noindent\textit{Generic social background.}
\begin{Prompt}
You supervise a pair of inventory assistants who work in the emergency-supply office.

You report through the hospital's standard management structure to its operations director.

You join managers from other departments at the hospital's weekly administrative planning meetings.
\end{Prompt}

\noindent\textit{Regional descriptive grounding.}
\begin{Prompt}
For two hospitals in the same regional emergency-supply network, you and the other manager serve as the designated supply contacts.

The network's procedure assesses conditions across participating hospitals during shortages.

The regional network's shortage procedure applies to your decision about managing Hospital A's designated batch.
\end{Prompt}

\noindent\textit{Local descriptive grounding.}
\begin{Prompt}
For hospitals in separate municipal emergency-supply systems, you and the other manager serve as the designated supply contacts.

Each system's procedure assesses conditions within its own hospital during shortages.

Hospital A's local shortage procedure applies to your decision about managing Hospital A's designated batch.
\end{Prompt}

\noindent\textit{Regional normative grounding.}
\begin{Prompt}
For two hospitals in the same regional emergency-supply network, you and the other manager serve as the designated supply contacts.

The network's procedure assesses conditions across participating hospitals during shortages.

When deciding how to manage Hospital A's designated batch, your role requires you to follow the regional network's shortage procedure.
\end{Prompt}

\noindent\textit{Local normative grounding.}
\begin{Prompt}
For hospitals in separate municipal emergency-supply systems, you and the other manager serve as the designated supply contacts.

Each system's procedure assesses conditions within its own hospital during shortages.

When deciding how to manage Hospital A's designated batch, your role requires you to follow Hospital A's local shortage procedure.
\end{Prompt}

\subsection{Execution, probability measurement, and analysis}
\label{app:experiment-measurement}

We used OpenAI's ``Responses'' API with temperature 0.7, reasoning effort \texttt{none}, a 64-token output limit, and zero retries. We used the returned token log-probabilities at the first output position and requested up to 20 candidates. The actions returned by the models were \texttt{TRANSFER} and \texttt{RE}, corresponding to the beginnings of \textsc{transfer} and \textsc{retain}. When both were present, we calculated transfer probability as $t/(t+r)$, where $t$ and $r$ denote their returned probability masses. We noticed after data collection that the provider had used a default top-$p$ of 0.98, causing one action candidate to be omitted in 339 of the 800 calls. For these calls, we bounded the transfer probability using the unreturned probability mass and used the interval midpoint for analysis. We ran a sensitivity analysis which showed that varying these estimates throughout their bounds changed aggregate measures by no more than 0.49 percentage points and preserved all context rankings within each model.

We average the ten call values for each combination of model, context condition, world, and surface format. For each world, surface sensitivity is the mean absolute difference between the canonical prose cell mean and each of the other three surface cell means. (We used canonical prose as the reference because the other surfaces were constructed as alternative renderings based on it.) The surface sensitivity measure averages the regional and local values. World responsiveness is the regional-minus-local difference in transfer probability, averaged across surfaces.

We obtain 95\% percentile intervals for repeated-call variation by resampling the ten calls with replacement within each cell and recomputing the measures 10,000 times.

\subsection{Additional results}
\label{app:experiment-additional-results}

Table~\ref{tab:experiment-full-results} reports the two plotted measures for all five context conditions.

\begin{table*}[t]
    \centering
    \small
    \caption{Selective robustness measures using ten calls per combination of model, context, world, and surface. Entries are point estimates with 95\% bootstrap intervals, in percentage points. Surface sensitivity is an average of the regional and local values.}
    \label{tab:experiment-full-results}
    \begin{tabular}{@{}llcc@{}}
        \toprule
        Model & Context
        & Mean surface sensitivity & World responsiveness \\
        \midrule
        Sol & Minimal
            & 34.1 [30.3, 40.3] & 33.9 [28.0, 40.2] \\
            & Neutral padding
            & 14.2 [10.7, 17.8] & 6.7 [1.2, 12.2] \\
            & Generic background
            & 27.9 [21.3, 33.7] & 51.1 [45.2, 56.6] \\
            & Descriptive grounding
            & 3.7 [1.8, 6.0] & 85.9 [82.4, 88.9] \\
            & Normative grounding
            & 3.9 [3.3, 4.6] & 92.0 [91.0, 92.9] \\
        \midrule
        Terra & Minimal
            & 19.8 [14.6, 26.9] & 38.6 [29.5, 47.9] \\
              & Neutral padding
            & 15.4 [13.6, 20.2] & 23.7 [17.5, 29.8] \\
              & Generic background
            & 30.6 [24.8, 38.1] & $-2.1$ [$-10.8$, 7.1] \\
              & Descriptive grounding
            & 13.2 [10.2, 18.3] & 57.2 [50.0, 64.1] \\
              & Normative grounding
            & 15.0 [14.2, 15.8] & 77.1 [76.0, 78.5] \\
        \bottomrule
    \end{tabular}
\end{table*}

Table~\ref{tab:experiment-cell-means} gives the means for the 80 cells.

\begin{table*}[t]
    \centering
    \scriptsize
    \caption{Mean transfer probability by model, context, world, and surface, in percent. C = canonical prose, B = bullets, K = key-value table, and A = alternative prose. Values below 0.05\% are shown as $<0.1$.}
    \label{tab:experiment-cell-means}
    \resizebox{\textwidth}{!}{%
    \begin{tabular}{@{}llrrrrrrrr@{}}
        \toprule
        & & \multicolumn{4}{c}{Regional world}
          & \multicolumn{4}{c}{Local world} \\
        \cmidrule(lr){3-6}\cmidrule(l){7-10}
        Model & Context & C & B & K & A & C & B & K & A \\
        \midrule
        Sol & Minimal
            & 37.8 & 41.0 & 94.8 & 88.9 & 15.3 & 29.8 & 1.5 & 80.5 \\
            & Neutral padding
            & 17.0 & 6.8 & 0.4 & 40.2 & 0.6 & 4.6 & 0.5 & 32.0 \\
            & Generic background
            & 15.8 & 48.7 & 87.8 & 66.8 & 1.6 & 8.3 & $<0.1$ & 4.7 \\
            & Descriptive grounding
            & 90.3 & 77.7 & 92.5 & 83.1 & $<0.1$ & 0.1 & $<0.1$ & 0.1 \\
            & Normative grounding
            & 96.6 & 77.6 & 99.2 & 94.5 & $<0.1$ & $<0.1$ & $<0.1$ & $<0.1$ \\
        \midrule
        Terra & Minimal
              & 99.8 & 51.4 & 98.8 & 100.0 & 63.0 & 69.4 & 46.4 & 16.8 \\
              & Neutral padding
              & 93.3 & 79.5 & 91.3 & 99.4 & 74.9 & 68.7 & 30.4 & 94.8 \\
              & Generic background
              & 41.5 & 18.8 & 78.9 & 92.6 & 46.0 & 62.6 & 37.9 & 93.7 \\
              & Descriptive grounding
              & 97.9 & 97.6 & 99.0 & 100.0 & 36.5 & 19.0 & 26.1 & 84.1 \\
              & Normative grounding
              & 100.0 & 99.9 & 100.0 & 100.0 & 0.9 & 0.4 & 0.2 & 89.9 \\
        \bottomrule
    \end{tabular}}
\end{table*}

The results in Table \ref{tab:experiment-full-results} show that grounding improves both measures relative to non-grounded conditions, but to different degrees across models. For Sol, moving from generic background to descriptive grounding reduces mean surface sensitivity from 27.9 to 3.7 percentage points and increases world responsiveness from 51.1 to 85.9 points; normative grounding further increases responsiveness to 92.0 and leaves surface sensitivity nearly the same at 3.9. For Terra, descriptive grounding reduces surface sensitivity from 30.6 to 13.2 points and raises world responsiveness from $-2.1$ to 57.2; normative grounding produces nearly the same surface sensitivity at 15.0 points, while increasing responsiveness to 77.1 points.

The cell means in Table~\ref{tab:experiment-cell-means} also reveal an important difference behind these aggregate results. Across Sol's descriptive and normative conditions, regional transfer probabilities range from 77.6\% to 99.2\%, while every local value is no more than 0.1\%. Terra's grounded regional cells similarly remain near ceiling, from 97.6\% to 100.0\%, but its local cells vary sharply across some surfaces. In the local descriptive condition, alternative prose yields 84.1\% transfer, compared with 19.0\% to 36.5\% for the other surfaces. In the local normative condition, the corresponding values are 89.9\% versus 0.2\% to 0.9\%. Thus, Terra's remaining surface variation under grounding is concentrated largely in the local alternative-prose cells.



\end{document}